\documentclass[12pt]{iopart} 	%preprint

\def\schaal{0.9} % Adjust parameter to set the scalefactor for scalable graphics

\usepackage{graphicx}
\usepackage{bm}

\begin{document}

\title[Conductance anisotropy and linear magnetoresistance in La${_{2-x}}$Sr${_x}$CuO${_4}$ thin films]{Conductance anisotropy and linear magnetoresistance in La$\bm{_{2-x}}$Sr$\bm{_x}$CuO$\bm{_4}$ thin films}

\author{M van Zalk$^1$, A Brinkman$^1$ and H Hilgenkamp$^{1,2}$}

\address{$^1$ Faculty of Science and Technology and MESA+ Institute for Nanotechnology, University of Twente, 7500 AE Enschede, The Netherlands}
\address{$^2$ Leiden Institute of Physics, Leiden University, 2300 RA Leiden, The Netherlands}
\ead{a.brinkman@utwente.nl}

\begin{abstract}
We have performed a detailed study of conductance anisotropy and magnetoresistance (MR) of La$_{2-x}$Sr$_x$CuO$_4$ (LSCO) thin films (0.10 $<x<$ 0.25). These two observables are promising for the detection of stripes. Subtle features of the conductance anisotropy are revealed by measuring the transverse resistance $R_{xy}$ in zero magnetic field. It is demonstrated that the sign of $R_{xy}$ depends on the orientation of the LSCO Hall bar with respect to the terrace structure of the substrate. Unit-cell-high substrate step edges must therefore be a dominant nucleation source for antiphase boundaries during film growth. We show that the measurement of $R_{xy}$ is sensitive enough to detect the cubic-tetragonal phase transition of the STO substrate at 105~K. The MR of LSCO thin films  shows for 0.10 $<x<$ 0.25 a non-monotonic temperature dependence, resulting from the onset of a linear term in the MR above 90~K. We show that the linear MR scales with the Hall resistivity as $\Delta \rho/\rho_0 \propto |\rho_{xy}(B)|$, with the constant of proportionality independent of temperature. Such scaling suggests that the linear MR originates from current distortions induced by structural or electronic inhomogeneities. The possible role of stripes for both the MR and the conductance anisotropy is discussed throughout the paper.
\end{abstract}

\pacs{73.50.Jt 74.25.fc 74.72.-h}

\submitto{\JPCM}

\maketitle

\section{Introduction \label{intro}}
Strong electron correlations lead to a wide variety of exceptional phenomena. In high-$T_\mathrm c$ superconductors strong correlations play an important role. For instance, they induce the Mott insulating state in the undoped parent compounds, despite the fact that the electronic bands are half filled. A rich phase diagram appears upon the introduction of charge carriers in these Mott insulators. Under certain circumstances charge carriers spontaneously order along lines, called stripes, which separate undoped antiferromagnetic (AF) regions \cite{Zaanen1989cmd}. Diffraction experiments have demonstrated (static) charge and spin modulation in La$_{1.6-x}$Nd$_{0.4}$Sr$_x$CuO$_4$ (Nd-LSCO) \cite{tranquada1995esc} and La$_{2-x}$Ba$_x$CuO$_4$ (LBCO) \cite{Fujita2002cbc,abbamonte2005smm},
leaving no doubt that stripes exist in these compounds. Two conditions need to be satisfied for stripe ordering to occur: (1) A doping near $x$ = 1/8, corresponding to a filling factor of 1/2 for Cu sites along the stripe. This condition relates stripes to the 1/8 anomaly \cite{Moodenbaugh1988spl}, a strong suppression of superconductivity at this doping. (2) A structural phase transition from the low-temperature orthorhombic (LTO) to the low-temperature tetragonal (LTT) phase, which is believed to provide a pinning potential for stripes, through the specific rotations of oxygen octahedra surrounding the Cu atoms \cite{tranquada1995esc}.

In a wider variety of compounds, among which La$_{2-x}$Sr$_x$CuO$_4$ (LSCO), incommensurate spin ordering is observed but no evidence for charge ordering \cite{Yamada1998dds,Lee1999nss,Arai1999isp,mook1999charge}. In LSCO such spin ordering can be observed throughout the doping range $x$ = 0.02--0.25 \cite{Yamada1998dds}. Peaks in neutron diffraction data (either at zero or finite energy) resemble these due to stripes and it is therefore reasonable to propose the presence of a fluctuating stripe phase when condition (1) and (2) for a static stripe phase are not fulfilled \cite{Kivelson2003hdf}.

While static stripe ordering in Nd-LSCO and LBCO has a pronounced effect on superconducting and transport properties, such as $T_\mathrm c$, the thermopower and the Hall coefficient $R_\mathrm H$ \cite{Nakamura1992atp,Hucker1998cso,Adachi2001cgt}, the consequences of fluctuating stripes\slash incommensurate spin correlations in LSCO and YBCO remain elusive. An interesting hypothesis is that \emph{if} fluctuating stripes are conducting \cite{kivelson1998elc} and fluctuate along some preferential direction, an anisotropy occurs in the macroscopic conductivity of the host material.  Ando \emph{et al.}~\cite{Ando2002era} have investigated conductance anisotropy in LSCO in the lightly hole-doped ($x$ = 0.02--0.04) region and in underdoped YBCO, finding the lowest resistance in the direction along the spin stripes. In addition to conductance anisotropy, several other fingerprints of stripes have been investigated. Anisotropic magnetoresistance (MR) was reported for underdoped YBCO and related to stripes \cite{Ando1999maa}.  Lavrov \emph{et al.}~\cite{Lavrov2003nsc} have searched for nonlinear current-voltage effects related to stripe motion induced by applied electric fields. Their negative result implies that if charged stripes exist in thin films, they should be pinned strongly.

In our work we proceed to investigate conductance anisotropy in LSCO thin films (0.10 $<x<$ 0.25) structured into Hall bridges oriented in various directions with respect to the LSCO Cu-O-Cu direction with 5$^\circ$ resolution. Furthermore, we investigate the transverse in-plane ($I$\,$\perp$\,$B$, $B$\,$\parallel$\,$c$) MR, motivated by the observation of linear transverse MR in LSCO single crystals for doping $x=$ 0.12--0.13 by Kimura \emph{et al.}~\cite{Kimura1996ipo}, which might well be a signature of a fluctuating stripe phase. We observe a sensitivity of the conductance anisotropy for lattice symmetry and we find indications for inhomogeneity on a small length scale. We carefully consider whether these could be due to the presence of stripes, discussing alternative explanations as well. In particular, we discuss the role of structural antiphase boundaries, which will be shown to be nucleated from substrate terrace edges.

\section{Experimental details\label{sec:exp}}

\begin{figure}
\centering
\includegraphics[scale=\schaal]{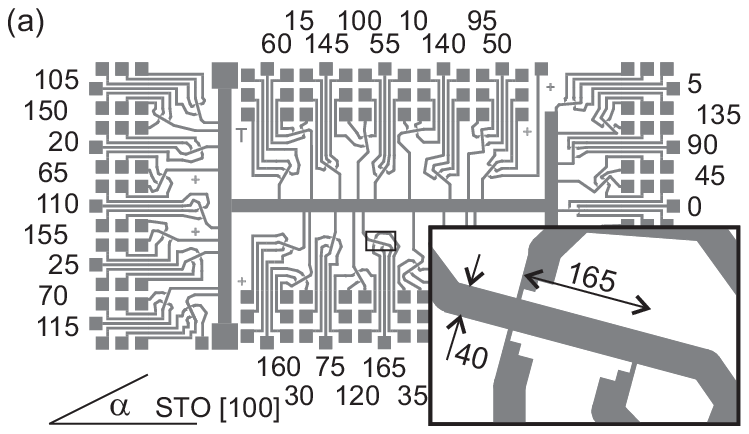}
\includegraphics[scale=\schaal]{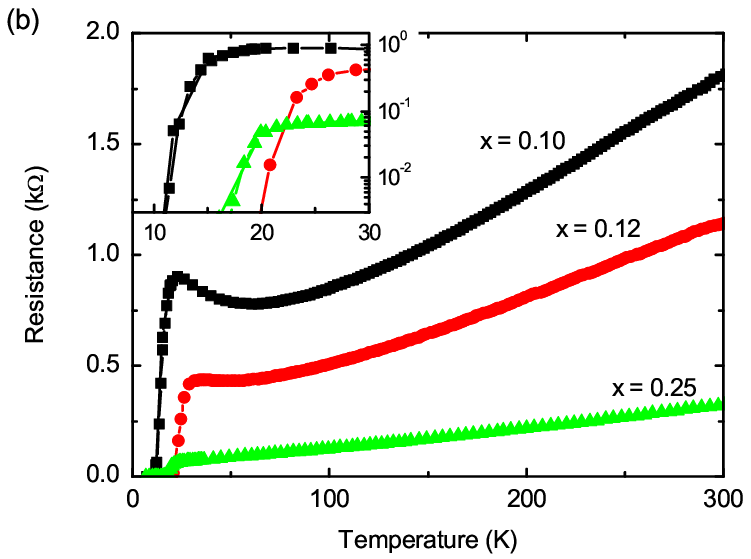}
\caption{\label{FIG1} (a) Sample structure consisting of 36 LSCO Hall bars (one of them shown in the inset) covering $\alpha =$ 0--175$^\circ$ with 5$^\circ$ resolution. Bonding pads and wiring leads are covered by Ti/Au. The STO [100] axis aligns with the long side of the sample. Hall bar dimensions are shown in the inset (in $\mu$m). (b) $R(T)$ curves for three different LSCO compositions.}
\end{figure}

LSCO thin films (thicknesses $d$ in the range 30--60 nm) were grown by pulsed laser ablation from sintered LSCO targets on SrTiO$_3$ (001) (STO), (La$_{0.3}$Sr$_{0.7}$)(Al$_{0.65}$Ta$_{0.35}$)O$_3$ (100) (LSAT), and NdGaO$_3$ (110) (NGO) substrates.  All STO substrates except one were chemically etched \cite{koster1998qis}. NGO and STO substrates were annealed for at least two hours at 950~$^\circ$C in an oxygen environment, LSAT substrates for 10 hrs at 1050~$^\circ$C. Atomic force microscopy (AFM) confirmed atomically flat substrate surfaces with unit-cell-height substrate steps. The miscut angle typically was 0.1--0.2$^\circ$.

Films were deposited in 0.13~mbar oxygen at a temperature of 700~$^\circ$C. The laser fluence was 1.2~J\,cm$^{-2}$. The film growth was monitored by reflective high-energy electron diffraction, which showed intensity oscillations, indicative for layer-by-layer growth. The thin films were annealed for 15 min at the deposition pressure and temperature, after which the oxygen pressure was increased to 1~atm, in which the films were annealed 15 min at 600~$^\circ$C, 30 min at 450~$^\circ$C and subsequently cooled down to room temperature.  

$c$-Axis oriented epitaxial growth was confirmed by x-ray diffraction. Lattice mismatches result in tensile strain values of 3.2\%, 2.4\%, and 2.0\% for STO, LSAT, and NGO, respectively.

Hall bars in various orientations [figure \ref{FIG1}(a)] were defined by photolithography and Ar-ion milling. The STO [100] axis aligns with the long side of the sample. For each experiment, insulating behavior of the substrate was confirmed. Electrical contacts were made by wire bonding to sputtered Ti/Au contact pads, defined by lift-off. Resistance and Hall measurements were performed in a commercial cryostat (Quantum Design, PPMS) with magnetic fields applied perpendicular to the thin film. Resistance measurements were independent of applied current (typically 1--100 $\mu$A) and Hall measurements were linear over the entire magnetic field range ($B=$ -9~T to +9~T). No significant changes in resistivity $\rho$ or $T_\mathrm c$ were observed as a result of thermal cycling.

Figure \ref{FIG1}(b) shows $R(T)$ plots for samples with different Sr contents. We verified that the target stoichiometry ($x$ = 0.10, 0.12, and 0.25) was transferred 1:1 to the thin film by comparing Hall coefficients obtained for our thin films with bulk values \cite{Ando2004ehc}. For the compositions $x$ = 0.10 and $x$ = 0.12, the Hall angle $\rho/R_\mathrm H$ showed a $T^2$-dependence over 50--300~K, whereas for $x$ = 0.25 $\rho/R_\mathrm H$ linearly depends on temperature. This behavior is in perfect agreement with reported high-quality single-crystal and thin-film data on LSCO \cite{Xiao1992uhe,Hwang1994std,Ando2004ehc}.

\section{Results and Discussion}
\subsection{Conductance anisotropy \label{anisotropy}}

\begin{figure}
\centering
\includegraphics[scale=\schaal]{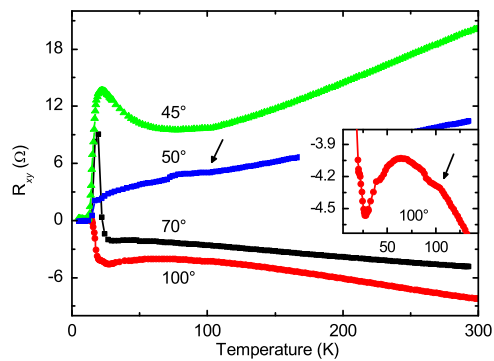}% Here is how to import EPS art
\caption{\label{FIG2} Conductance anisotropy measured by the transverse resistance for an LSCO thin film ($x=0.12$) for different orientations $\alpha$. Arrows indicate anomalies which will be discussed in Sec.~\ref{anomalies}. }
\end{figure}

\begin{figure}
\centering
\includegraphics[scale=\schaal]{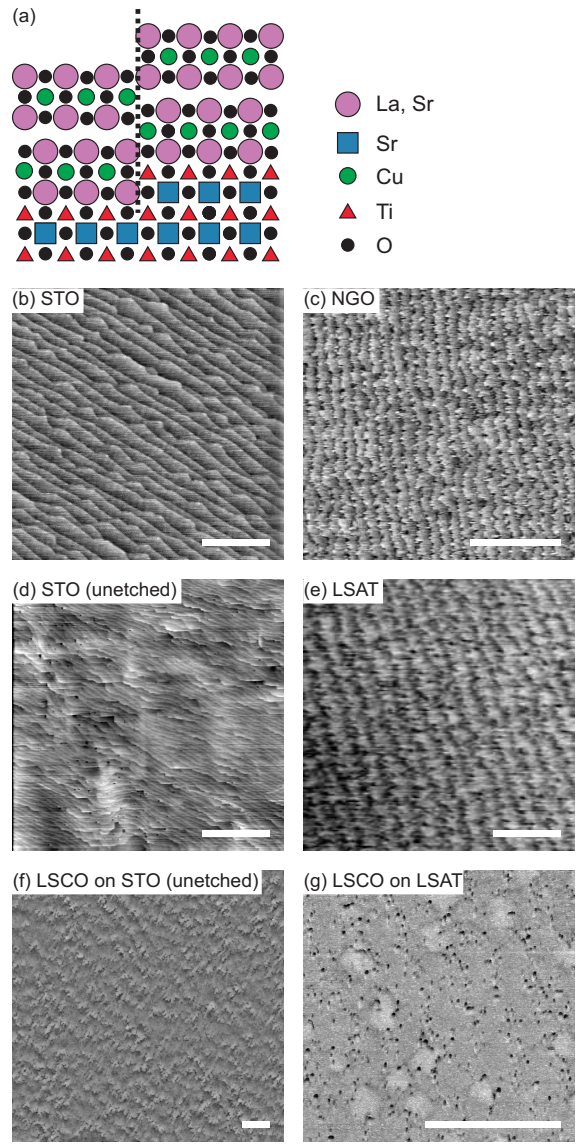}% Here is how to import EPS art
\caption{\label{FIG3} (a) Schematical representation of a step-edge induced antiphase boundary (dashed line) in LSCO on STO. Substrate (b--e) and LSCO thin film (f,g) surfaces. Films shown in (f) and (g) were grown on substrates in (d) and (e), respectively. Scale bars denote 1~$\mu$m.}
\end{figure}

\begin{figure}
\centering
\includegraphics[scale=\schaal]{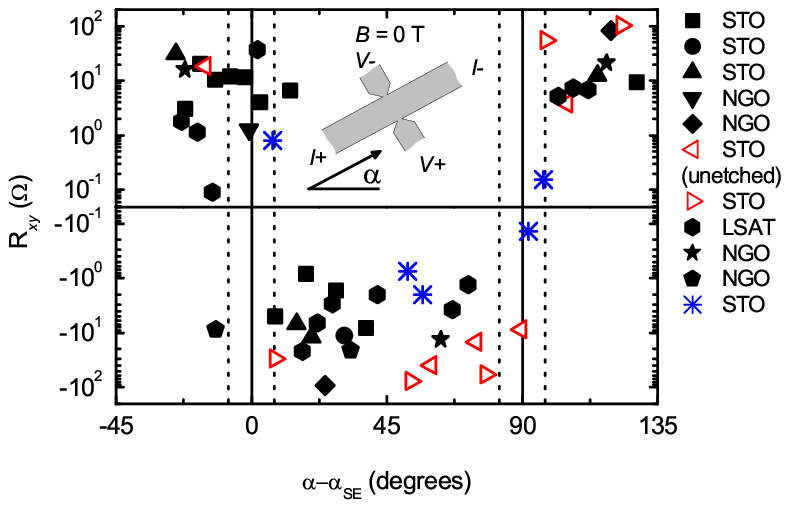}% Here is how to import EPS art
\caption{\label{FIG4} Room temperature transverse resistance ($B$ = 0~T) versus Hall bar orientation with respect to step-edge direction for different substrates and $x$ = 0.10 (red, open symbols), $x$ = 0.12 (black, solid), and $x$ = 0.25 (blue crosses). $\alpha_\mathrm{se}$ varies randomly between 10 and 140$^\circ$. }
\end{figure}

Conductance anisotropy, a smoking gun for the presence of fluctuating stripes \cite{Ando2002era}, is most effectively examined by measuring the transverse resistance $R_{xy} = U_y/I_x(\alpha)$ ($x$ and $y$ orthogonal directions) for $B$ = 0~T, since the angle-dependence of the longitudinal resistance $R(\alpha)$ is easily affected by small inhomogeneities in the sample. We find a relatively large signal for $R_{xy}$ for all substrates and doping (figure \ref{FIG2}), which cannot be attributed to misalignment of the voltage contacts, given the resolution of the applied photolithography technique. 
 
One possible explanation for the anisotropy is the stepped substrate surface, which might induce structural antiphase boundaries in the film. Such antiphase boundaries have experimentally been observed in YBa$_2$Cu$_3$O$_7$ on STO using high-resolution electron microscopy \cite{Wen1993ssy}. Figure \ref{FIG3}(a) shows schematically how an antiphase boundary would look like for LSCO. The CuO$_2$-planes are interrupted at the structural antiphase boundary. Typical surfaces of our substrates as measured by AFM are shown in figures \ref{FIG3}(b--e). In (f,g) it can be seen that the film surface reflects the morphology of the substrate. We have determined the step-edge orientation $\alpha_\mathrm{se}$ of all our substrates from AFM data obtained before deposition of the LSCO thin films.

Figure \ref{FIG4} shows that the sign of $R_{xy}$ can be predicted with high certainty from the orientation of the Hall bar with respect to the step-edge orientation ($\alpha - \alpha_\mathrm{se}$) for all our samples. This provides evidence that antiphase boundaries in LSCO thin films are dominantly nucleated from substrate step edges. The large spread in $R_{xy}$ reflects the randomness exhibited by step edges. For $x$ = 0.10--0.12, we estimate an antiphase-boundary resistivity of $\rho_\mathrm{AB} \approx 10^{-9}$~$\Omega$\,cm$^2$ at room temperature, which is in line with typical interface resistances involving high-$T_\mathrm c$ cuprates \cite{Beck1996lbg,Hilgenkamp2002gbh}.
From a typical critical current density value ($J_\mathrm c \approx$~10$^6$~A\,cm$^{-2}$) we estimate an $I_\mathrm c R_\mathrm n$ product of about 1~mV, which is a reasonable value \cite{Hilgenkamp2002gbh}.
For $x$ = 0.25, $\rho_\mathrm{AB}$ is about 10 times smaller, which is in agreement with an expected decrease in thickness of the depletion region \cite{Hilgenkamp2002gbh}.

\subsection{Conductance anisotropy anomaly at 105~K \label{anomalies}}

\begin{figure}
\centering
\includegraphics[scale=\schaal]{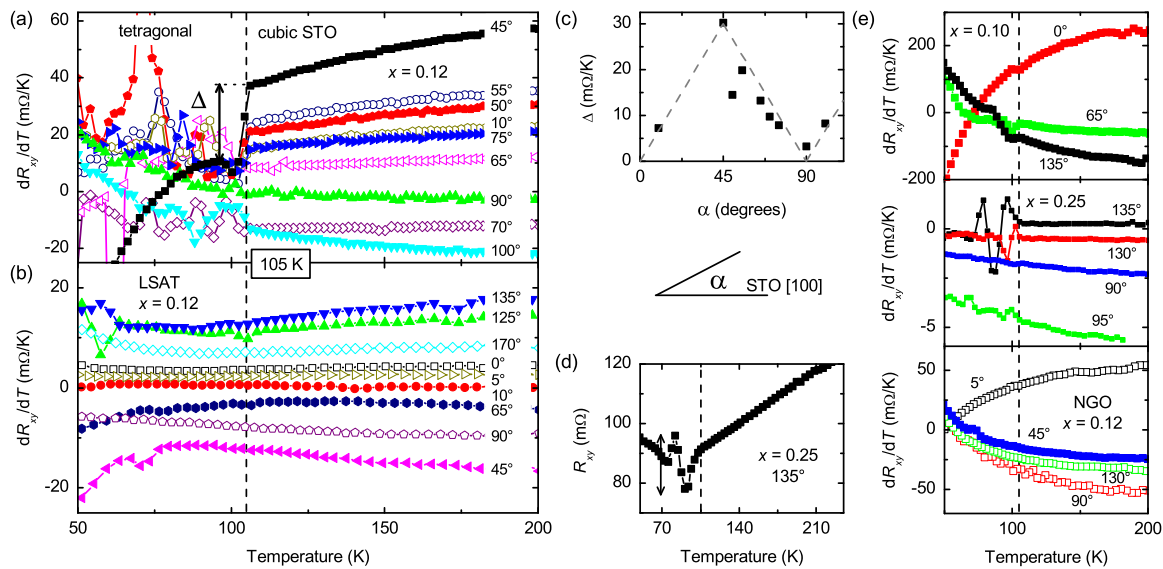}% Here is how to import EPS art
\caption{\label{FIG5} (a) Numerical derivative of the transverse resistance $R_{xy}$ showing a clear jump $\Delta$ at 105~K, coincident with the cubic-tetragonal transition in the STO substrate at 105~K. (b) These effects are absent on LSAT, providing evidence that $\Delta$ is indeed related to the STO phase transition. On LSAT weak instabilities are found in the range 60--80~K predominantly for orientations close to the Cu-Cu direction (the curves showing such an instability are plotted by solid symbols). (c) Orientational dependence of $\Delta$, measured at 105 K for LSCO on STO (d) $R_{xy}$ for $x=0.25$, in which the anomaly can be observed without differentiation because of the small value of the antiphase boundary induced background. The arrow denotes the  resistance change estimated from the stress developing in the LSCO layer upon the the STO phase transition, using the pressure-dependent resistivity data from Nakamura \emph{et al.}~\cite{Nakamura1999tpl}. (e) As (a) and (b) but for $x=0.10$ and $x = 0.25$ (both on STO) and for $x= 0.12$ on NGO. The latter shows an instability at $T=78$~K for $\alpha = 45^\circ$.}
\end{figure}

In figure \ref{FIG2}, anomalies can be observed in $R_{xy}(T)$ at 105 K. These are most clearly revealed upon numerical differentiation. Figure \ref{FIG5} shows that discontinuities in $\mathrm d R_{xy}/\mathrm d T$ are present for all doping, however only for STO substrates. By defining $\Delta \equiv  (\mathrm d R_{xy}/\mathrm d T)_{T\downarrow \mathrm{105 K}} - (\mathrm d R_{xy}/\mathrm d T)_{T\uparrow \mathrm{105 K}}$ we demonstrate in figure \ref{FIG5}(c) that $\Delta$ depends on the Hall bar orientation $\alpha$. The largest $\Delta$ is observed for $\alpha = 45^\circ$, whereas for $\alpha = 90^\circ$ $\Delta$ hardly exceeds the noise level. We do not observe anomalies in the longitudinal resistance.

The sudden change in $\mathrm d R_{xy}/\mathrm d T$ at 105~K coincides with a cubic-tetragonal phase transition in STO \cite{Courtens1972bsp}. The fact that such behavior is only observed for STO substrates proves that it is in fact induced by this structural transition. Noise in $\mathrm d R_{xy}/\mathrm d T$ below 105~K can then be attributed to rearrangement of domains in the substrate, since the $c$ axis can align along three orthogonal directions. The deviation from the cubic unit cell in the tetragonal phase ($T<105$~K) is small ($c/a$ = 1.00056 at 56~K \cite{Lytle1964xrd}). Since the LSCO film is epitaxially connected to the substrate, we expect the change of the substrate's lattice to be fully passed on to the LSCO film.

The lattice parameter changes associated with the LTO-LTT transition in LBCO ($a_\mathrm {LTT}/a_\mathrm {LTO}$ = 1.0017 and $b_\mathrm{LTO}/a_\mathrm{LTO}$ = 1.0036 \cite{Katano1993css}) are a few times larger than the structural changes induced by the STO. Yet, for LBCO these small modifications represent a significant change in the tilting direction of the oxygen octahedra, providing the necessary pinning potential to stabilize a static stripe phase \cite{tranquada1995esc}. Pinning of the fluctuating stripe phase present in LSCO as a result of the induced lattice asymmetry by the STO phase transition would naturally lead to the observed conductance anisotropy change.  There are however a few difficulties with this stripe pinning scenario. First, one might expect a stronger doping dependence, as a static stripe phase appears in single crystals of Nd-LSCO and LBCO only around $x = 1/8$. Second, the appearance of static stripes in these compounds coincides with discontinuities in transport properties, in particular in $R_\mathrm H$ \cite{Nakamura1992atp, Adachi2001cgt}. We do not observe any peculiarity in $R_\mathrm H$ around 105~K.

Transport properties in LSCO and other high-$T_\mathrm c$ compounds are generally sensitive to applied pressure, pointing toward a delicate dependence of electronic structure on crystal structure \cite{Yamada1992pes,Nakamura1999tpl}. The observed conductance anisotropy anomalies might therefore be a manifestation of pressure effects on transport properties.  We estimate the stress developing in the LSCO layer due to the strain change at 105~K from the Young's modulus of 10$^{11}$--10$^{12}$ Pa \cite{Nakamura1999tpl,Sarrao1994cem} to be 0.06--0.6 GPa. Using data from Nakamura \emph{et al.}~\cite{Nakamura1999tpl} we estimate for $x$ = 0.25 at 105~K a maximum resistivity change induced by such stress of 10$^{-7}$~$\Omega$\,cm, leading to $\Delta R_{xy} \approx 20$~m$\Omega$ for our structure. This value compares well to the measured $\Delta R_{xy}$ for this doping; see the arrow in figure \ref{FIG5}(d). For lower $x$, the pressure dependence of LSCO is stronger and $\Delta R_{xy}$ will likely be larger. This is consistent with our observations, although a quantitative comparison is difficult because antiphase boundaries induce a stronger background in $R_{xy}$. The expected stress effect in the longitudinal resistance ($\Delta R \approx 4 \Delta R_{xy}$) is smaller than the noise we measure in $R$, which explains why we do not observe anomalies in $R$. Only the differential measurement of $R_{xy}$ is sensitive enough to reveal the STO cubic-tetragonal phase transition through a resistivity measurement.

For NGO, no structural phase transitions are reported in the temperature range 50--200~K \cite{Senyshyn2004tep}. For LSAT there might be small distortion from cubic symmetry at and below 150~K \cite{chakoumakos1998tel}. We do not observe a transition near 150~K in \ref{FIG5}(b). Weak fluctuations for $T>$~150~K could be traced to variation in the temperature sweep rate. Both on LSAT and NGO [figures \ref{FIG5}(b) and (e), bottom panel] we observe instabilities in $\mathrm d R_{xy}/\mathrm d T$ in the temperature range 60--80~K, predominantly for Hall bar orientations close to $\alpha$ = 45$^\circ$ and 135$^\circ$. These instabilities are only observed when cooling down, and not for increasing temperature, unlike the effects on STO. Although speculative, they could be explained by a structural transition in the film, rather than in the substrate. Perhaps the high-temperature tetragonal (HTT) structure is sufficiently clamped by the substrate to reduce the transition to the low-temperature orthorhombic (LTO) phase to 60--80~K.

\subsection{Magnetoresistance}

\begin{figure}
\centering
\includegraphics[scale=\schaal]{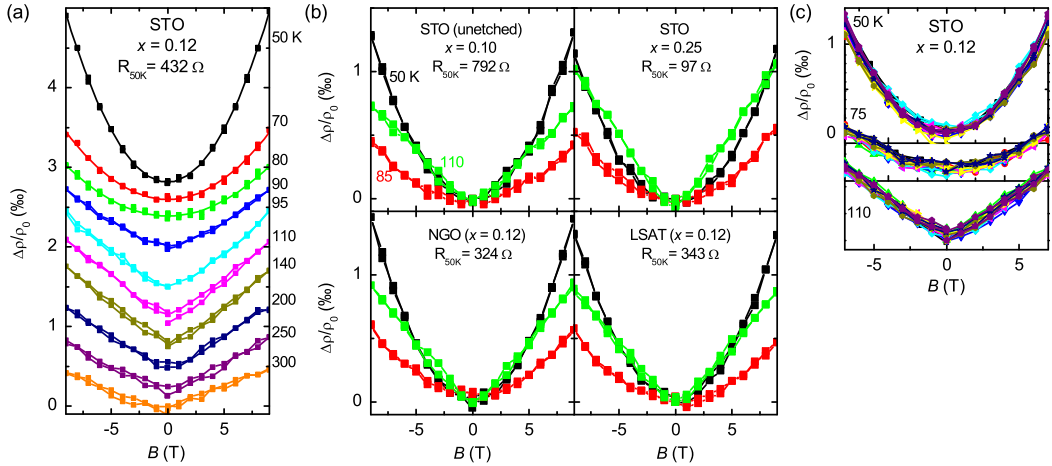}% Here is how to import EPS art
\caption{\label{FIG6} (a) Temperature evolution of MR in LSCO on STO ($x$ = 0.12). Solid lines are parabolic fits to data for 50, 70, and 80~K. At 90~K, a crossover to linear MR is observed. (b) The onset of linear MR between 85 and 110~K is observed for various substrates and $x$. The zero-field resistances at 50~K are shown in the graphs. (c) MR for several Hall bars on two different samples. We do not observe sample-to-sample variations. }
\end{figure}

Magnetoresistive properties of LSCO and high-$T_\mathrm c$ cuprates in general have been investigated widely, both in the superconducting ($T<T_\mathrm c$) regime \cite{suzuki1991rtm,Ando1995ldb,Xiang2009fsa}, as in the normal state \cite{Kimura1996ipo,Balakirev1998oml,Harris1995vkr,Ando2003aml,Vanacken2005hfm}. Many studies have focused on the violation of Kohler's rule \cite{Harris1995vkr}, anisotropy of MR in relation to stripes \cite{Ando2003aml}, and high magnetic fields \cite{Vanacken2005hfm}. Most work has been done with single crystals. Here we show that the low-field magnetoresistance $\Delta \rho/\rho_0$ of LSCO thin films shows intriguing non-monotonic behavior as function of temperature with a crossover from quadratic to linear MR at 90~K. Such behavior [figure \ref{FIG6}(a,b)] is observed for all doping values and all substrates that were used for this research. Literature reports \cite{Balakirev1998oml} quadratic MR without linear component for much thicker LSCO films on LaSrAlO$_3$, which puts the LSCO under \emph{compressive} strain (with a moderate lattice mismatch of 0.5\%).

The linear MR ($T>$ 90~K) in our thin films is weakly dependent on $x$ and substrate type, and comparable in magnitude to linear MR reported in single crystals ($x$ = 0.12--0.13) by Kimura \emph{et al.}~\cite{Kimura1996ipo}. In both cases, linear MR weakly decreases with increasing temperature over 90--300~K. The quadratic component ($T<$ 90~K) in our data is suppressed rapidly between 50 and 85~K. This behavior is similar for single crystals. The crossover that we observe at 90~K, might therefore be interpreted as a sudden onset of a linear term above 90~K in combination with a gradual suppression of quadratic MR with increasing temperature. Interestingly, the linear MR appearing in single crystals ($x$ = 0.12--0.13) is present down to 50~K, and as a result, MR decreases monotonically with temperature. 

\begin{figure}
\centering
\includegraphics[scale=\schaal]{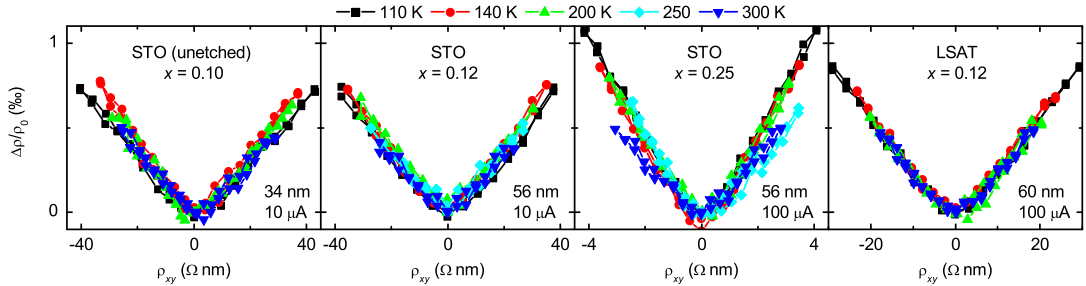}% Here is how to import EPS art
\caption{\label{FIG7} The magnetoresistance plotted versus the Hall resistivity. We observe scaling as $\Delta \rho/\rho_0 \propto |\rho_{xy}(B)|$, with the constant of proportionality independent of temperature. Film thicknesses and measurement currents are specified in the graphs.}
\end{figure}

The doping dependence of linear MR observed in single crystals strongly suggests a relation to the $1/8$ anomaly. Kimura \emph{et al.}~\cite{Kimura1996ipo} propose it to result from magnetic field enhanced fluctuations towards the stripe phase. An alternative explanation in terms of a van Hove singularity crossing the Fermi energy at $x \approx$ 0.13 has become obsolete by more recent angle-resolved photoemission spectroscopy \cite{Ino2002dde}. The absence of doping dependence for linear MR in thin films makes an explanation in terms of fluctuating stripes less likely. Dynamical incommensurate spin correlations, which might be indicative for fluctuating stripes, have been observed for the entire doping range $x$ = 0.05--0.25 \cite{Yamada1998dds}. Nevertheless, one would expect singular behavior near $x=1/8$ because many properties related to spin fluctuations are anomalous at this doping, such as the peak width in inelastic neutron scattering data \cite{Yamada1998dds} and the magnetic correlation length \cite{Kimura1999nss}.  Moreover, it is not clear whether relatively low magnetic fields can affect stripes since $B> 70$~T is required to meet the energy scale typical for dynamical spin correlations ($\mu_\mathrm B B> 4$~meV). Linearity of MR holds down to roughly 1~T both in single crystals as in thin films. Lastly, if linear MR in LSCO would be related to stripes, it is unclear why thin films would have a different doping dependence than single crystals. The same reasoning holds for all intrinsic explanations for linear MR, e.g., spin-mediated mechanisms. An obvious difference between single crystals and thin films is the presence of antiphase boundaries in the latter, as discussed in section \ref{anisotropy}.

Recently, large linear and non-saturating MR was reported for non-magnetic silver chalcogenides \cite{Xu1997lmn,Husmann2002mgs} and InSb \cite{Hu2008cqr}. It was argued by Parish and Littlewood \cite{Parish2003nmh} that the observed low-field MR can arise from sample inhomogeneity, present in the form of nanowires of excess Ag in the silver chalcogenides and Sb droplets in InSb polycrystals \cite{Hu2008cqr}. The linearity originates from a misalignment between applied voltage and current paths, which results in the mixing of Hall and longitudinal voltages. Since our samples exhibit mobilities of $\mu \approx$ 5--7 cm$^2$/Vs (at 50~K) all our measurements are taken in the low-field ($\mu B \ll 1$) regime. For Ag$_{2+\delta}$Se it was shown \cite{Husmann2002mgs} that the magnetoresistance follows a modified Kohler's rule: $b(T)\Delta \rho/\rho_0 = f(\rho_{xy}/\rho_0)$, with $B$ and the carrier density $n$ entering implicitly through $\rho_{xy}/d = R_{xy}(B,n)$. In our case we find a surprisingly simple non-Kohler type scaling: $\Delta \rho/\rho_0 \propto |\rho_{xy}(B)|$, with the constant of proportionality being independent of temperature; see figure \ref{FIG7}. This suggests the linear term in the MR has the same origin as the Hall resistivity. Clearly the mixing of the Hall and longitudinal resistances would provide a straightforward explanation for this behavior. 

One might wonder why linear MR for $x$ = 0.25 is slightly larger in magnitude than linear MR for $x$ = 0.10 and $x$ = 0.12, despite the fact that the antiphase-boundary resistivity is significantly smaller for $x$ = 0.25. It should be noted that also the resistivity of LSCO itself is much smaller for this doping and we expect the ratio between the two to determine the strength of linear MR. The inhomogeneity scenario also provides a natural explanation for absence of linear MR in much thicker LSCO films \cite{Balakirev1998oml}: the effects of the structural antiphase-boundaries might be washed out toward the thin film surface by the introduction of other types of defects, giving rise to more isotropic disorder. Moreover, the Hall voltage is smaller for thicker films, as it is inversely proportional to the film thickness.
 
Some questions remain concerning the inhomogeneity scenario. First, it is unclear why linear MR vanishes below 90 K. Both the longitudinal and Hall resistivity do not show apparent changes of behavior around 90 K. Second, we do not observe a strong sample-to-sample variation in the magnitude of linear MR [see figure \ref{FIG6}(c)], which might be expected if inhomogeneity is the underlying cause. Third, there is no dependence of linear MR on the Hall bar angle ($\alpha - \alpha_\mathrm{se}$). The answers to the last two questions may reside in the exact identification of inhomogeneity in our samples. Perfectly straight and parallel antiphase boundaries, with homogeneous $\rho_\mathrm{AB}$ might not give rise to linear MR as the current would be homogeneously distributed and flowing parallel to the Hall bar. Deviations from this perfect picture more likely cause linear MR and do not necessarily depend on $\alpha - \alpha_\mathrm{se}$. The small length scales of such imperfections might provide enough averaging to prevent sample-to-sample variations. Numerical calculations will have to corroborate the proposed scenario. If the mechanism would fail to account for our observations, an electronic origin (e.g. stripes) of linear MR will have to be reconsidered.

\section{Conclusion}
The transverse resistance $R_{xy}$ in zero magnetic field, usually background in a Hall measurement, provides valuable information about the microstructure of the material under study. We have used it to demonstrate that unit-cell-high substrate step edges are the dominant source of structural antiphase boundaries in LSCO thin films. The antiphase boundary resistivity was estimated to be $\rho_\mathrm{AB} \approx 10^{-9}$~$\Omega$\,cm$^2$ (room temperature). In addition, we show that for LSCO $R_{xy}$ can reveal structural phase transitions of the substrate on which the films are grown. Such transitions are usually difficult to detect and require advanced spectroscopic analysis equipment.  

For the detection of stripes, conductance anisotropy is an important observable. We have shown that in LSCO thin films conductance anisotropy is dominantly caused by antiphase boundaries, which mask possible stripe effects. Future experiments in this direction will therefore require substrates with an extremely small vicinal angle, and Hall bars at sub-micron scale. 

The silver chalcogenides have recently attracted interest because of their non-saturating linear MR, which make them suitable for use as magnetic field sensor \cite{Xu1997lmn,Husmann2002mgs}. The MR is linear down to surprisingly low magnetic fields in these materials. This has been explained by the presence of disorder, giving rise to the mixing of longitudinal and Hall resistances \cite{Parish2003nmh}. Our LSCO thin films show linear low-field MR in the entire doping range 0.10 $<x<$ 0.25. We have found the MR to scale with the Hall resistivity as $\Delta \rho/\rho_0 \propto |\rho_{xy}(B)|$ with a temperature-independent constant of proportionality. This suggests  the linear MR of LSCO thin films is related to disorder as well. Structural antiphase boundaries generated from substrate steps are a likely source of disorder. However, linear MR also appears in single crystals of LSCO, although in a narrower doping range ($x$ = 0.12--0.13) \cite{Kimura1996ipo}. It is unclear why these crystals in particular would contain many antiphase boundaries. If the presence of such defects can be excluded experimentally, linear MR must have a different origin, at least in single crystals. In that case it will be worth reconsidering the role of stripes, which might similarly deflect the current from the longitudinal direction, causing the mixing of longitudinal and Hall resistances.

\ack

We gratefully acknowledge Jan Zaanen for fruitful discussions. This work is financially supported by the Dutch Foundation for Fundamental Research on Matter (FOM), the Netherlands Organization for Scientific Research (NWO) through VIDI (A.B.) and VICI (H.H.) grants, and the NanoNed program.

\section*{References}
\bibliographystyle{iopart-num}
\bibliography{LSCO}
\end{document}